\documentstyle[11 pt,aas2pp4]{article}

\lefthead{Phillipps et al.}

\righthead{A Dwarf Population - Density Relation}

\begin{document}

\title{The Luminosity Distribution in Galaxy Clusters;\\
A Dwarf Population - Density Relation?}

\author{S. Phillipps}
\affil{Astrophysics Group, Department of Physics, University of Bristol,\\
Tyndall Avenue, Bristol BS8 1TL, U.K.}

\author{S.P. Driver and W.J. Couch}
\affil{School of Physics, University of New South Wales, \\
Sydney, NSW 2052, Australia}

\and

\author{R.M. Smith}
\affil{Department of Physics and Astronomy, University of Wales College of Cardiff,\\ 
PO Box 913, Cardiff, Wales, U.K.}

\begin{abstract}
Recent work suggests that rich clusters of galaxies commonly have large
populations of dwarf (ie. low luminosity) members, that is their luminosity
function (LF) turns up to a steep slope at the faint end. This population,
or more particularly the relative numbers of dwarfs to giants, appears
to be very similar for clusters of similar morphology, but may vary between 
cluster types. We have previously suggested that dwarfs may be more common 
in less compact, spiral rich clusters. Similarly we have found evidence for
population gradients across clusters, in that the dwarf population appears
more spatially extended.
In the present paper we summarise the current evidence and propose, in
analogy to the well-known morphology - density relation, that what we are
seeing is a dwarf population - density relation; dwarfs are more common
in lower density environments. 
Finally we discuss recent semi-analytic models of galaxy formation in the
hierarchical clustering picture, which may give clues as to the origin
of our proposed relation.
\end{abstract}


\section{Introduction}

Much recent work has been devoted to the question of the galaxy luminosity function
(LF) within rich clusters, particularly with regard to the faint end which has
become accessible to detailed study through various technical and 
observational improvements (see e.g., Driver et al. 1994; Biviano et al. 
1995; Bernstein et al. 1995; Mohr et al. 1996; 
Wilson et al. 1997; Smith, Driver \& Phillipps 1997 = Paper I; Trentham 1997a,b).

For the most part these studies concur that the LF becomes steep (Schechter (1976) slope $\alpha \leq - 1.5$) faintwards of about $M_{B} = -17.5$ 
or $M_{R} \simeq -19$ (for
$H_{0}$ = 50 km s$^{-1}$ Mpc$^{-1}$), and Paper I suggested that such a
dwarf rich population might be ubiquitous. In a subsequent paper (Driver, 
Couch \& Phillipps 1997a = Paper III) we have examined the
luminosity distribution in and across a variety of clusters, examining
the possible dependence of the dwarf population (in
particular the ratio of dwarfs to giants) on cluster type and position
within the cluster. In the
present paper we summarise the evidence to date for the (dis)similarity
of the dwarf population in different environments.

\section{The Dwarf Luminosity Function}

\subsection{Dwarfs in Rich Clusters}

Several papers (e.g., Driver et al. 1994; Paper I; Wilson et al. 1997)
have recently demonstrated remarkably similar dwarf populations in
a number of morphologically similar dense rich clusters like (and
including) Coma. This similarity appears not only in the faint end
slope of the LF, around $\alpha = -1.8$, but also in the point at
which the steep slope cuts in, $M_{R} \simeq -19$ (ie. about $M^{*} + 3.5$). 
The latter implies equal ratios
of dwarf to giant galaxy numbers in the different clusters.

However, there clearly do exist differences between some clusters. 
For example,
several of the clusters in the Paper III sample do not
show a conspicuous turn up at the faint end. Either these clusters
contain completely different types of dwarf galaxy population or, as we
suggest, the turn up occurs at fainter magnitudes (ie. the dwarf to
giant ratio, DGR, is smaller; we will define the DGR as in Paper III as the number of galaxies with $-16.5 \geq M_{R} \geq -19.5$
compared to those with $-19.5 \geq M_{R}$). 
This later upturn is, in principle, verifiable with yet deeper
photometry, though background contamination uncertainties eventually
dominate (see, e.g., Driver et al. 1997b = Paper II; Trentham 1997a).

The clusters in question are not distinguished by their richness, but
we can also check for morphological differences. As is well known,
structural and (giant galaxy) population characteristics are well correlated,
with, for example, dense regular clusters being of early Bautz-Morgan type
(dominated by cD galaxies) and having the highest fractions of giant
ellipticals. We can therefore choose to characterise the clusters by their
central (giant) galaxy number densities, for instance the number of galaxies brighter
than $M_{R} = -19.5$ within the central 1 Mpc$^{2}$ area. (An alternative
would be to use
Dressler's (1980) measure of the average number of near neighbours, see
Paper III). We find that the clusters with less
prominent dwarf populations (lower DGRs $\sim 1$)
are just those with the highest projected
galaxy densities (eg. A3888). The equivalent effect of earlier
B-M type clusters having less dwarfs (A3888 is B-M Type I-II)
was illustrated in Paper III (see also Lopez-Cruz et al. 1997). 
Earlier, Turner et al. (1993) had
noted that the rich but low density
cluster A3574, which is very spiral rich (Willmer et al. 1991), 
had a very high low surface brightness (LSB) dwarf to
giant ratio. This is now backed up by the observations (see Paper III)
of the clusters
like A204 which are dwarf rich (DGR $\sim 3$), have low central densities 
and late B-M types (A204 is B-M III). In addition,
it appears that the Virgo Cluster, an archetypal loose (moderately rich)
cluster, has a very large dwarf population (Binggeli, Sandage \& Tammann
1985), with many LSB dwarfs down to the sizes of Local Group dwarf spheroidals
(Schwartzenberg, Phillipps \& Parker 1996; Phillipps et al. 1998).

\subsection{Population Gradients}

It was already suggested by the results on A2554 in Paper I, that the dwarf
population was more spatially extended than that of the giants, ie. the
dwarf to giant ratio increased outwards. This type of population gradient 
has been confirmed by the results in Paper III. (We also consider this in more
detail elsewhere, Smith et al., in preparation = Paper IV, using observations
extending over larger areas). It is found, too, in Virgo
(Phillipps et al. 1998), where the dwarf LSBG population has almost constant
number density across the central areas while the giant density drops
by a factor $\sim 3$. For Coma, the apparent discrepancy
between, for example, 
the LF slopes of Bernstein et al. (1995), for the core, and Biviano
et al. (1995), for a larger area, could similarly be attributed to an
increase in the dwarf fraction in the outer parts (see also Thompson \& Gregory
1993, Karachentsev et al. 1995, and section 3 below).

\subsection{The Field}

The opposite extreme of environment to the elliptical rich
core of a dense cluster is, of course, the spiral rich `field', which
we can think of as made up of loose groups, and possibly the outskirts
of richer systems. Extending the above arguments we would therefore
expect to see a very large dwarf population in the field. This is
against the usual perception of the field LF at the faint end (e.g., Efstathiou,
Ellis \& Peterson 1988; Loveday et al. 1992), but evidence has begun to
accumulate that the field LF may indeed turn up at the faint end,
at a similar point to that seen in the clusters, around $M_{R} \simeq -19$
(e.g., Marzke, Huchra \& Geller 1994; Zucca et al. 1997; see also  
Driver \& Phillipps 1996). Very recently, considerations of the satellites of
nearby spiral galaxies have suggested that the field dwarf LF may be
just as steep as that in clusters (Morgan, Smith \& Phillipps 1998), or,
indeed, much steeper (Loveday 1997). A contributory factor to this revision
of the field dwarf LF is clearly the inclusion of LSBGs missed from
earlier surveys (Phillipps \& Driver 1995; Ferguson \& McGaugh 1995),
but it should be said that in the Local Group, where extremely low surface
brightness dwarfs are, in principle, detectable from their resolved stars,
the LF appears quite flat. From the relationship derived in Paper III, a
field LF with a faint end slope, say, $\alpha = -1.5$ would have
a DGR of 4 -- 5. This would be nicely consistent with the extrapolation 
of the trend 
seen in the clusters (see below) down to low (volume) densities. 

\section{A Population Density Relation}

The obvious synthesis of the above results is to posit a relationship
between the local galaxy density and the fraction of dwarfs (ie. the relative
amplitude of the dwarf LF). The inner, densest parts of rich clusters
would have the smallest fraction of dwarfs, while loose clusters and the
outer parts of regular clusters, where the density is lower, have high dwarf
fractions. This is illustrated in Figure 1, which is based on our
homogeneous rich cluster data from Papers I and III. 
It is particularly interesting
to note the clear overlap region, where regions of low density on the
outskirts of dense clusters (open squares) have the same DGRs as the
regions of the same density at the centres of looser clusters (solid squares).
The triangles show in slightly more detail the run of DGR with radius
(hence density) across an individual cluster, A2554.

The proposed relation
of course mimics the well known morphology - density relation (Dressler 
1980), wherein the central parts of rich clusters have the highest early type
galaxy fraction, this fraction then declining with decreasing local galaxy 
density.
Putting the two relations together, it would also imply that dwarfs 
preferentially occur in the same environments as spirals. This would be in 
agreement with a weaker clustering of low luminosity systems in general
(Loveday et al. 1995; see also Dominguez-Tenreiro et al.
1996) as well as for spirals compared to ellipticals
(Geller \& Davies 1976). Thuan et al. (1991) have previously discussed the
similar spatial distributions of dwarfs (in particular dwarf irregulars)
and larger late type systems. 

In Figure 2 we have added to our data (squares) values derived from the
work of other observers. A problem here is, of course, lack of homogeneity
due to different observed wavebands, different object detection
techniques and so forth.
Nevertheless, we can explore whether literature results support our results. 
Firstly, several points are shown for various surveys of Coma (hexagons).
These surveys (Thompson \& Gregory 1993, Lobo et al. 1997, Secker \& Harris 
1996 and
Trentham 1998) cover different areas and hence different mean projected
densities. All these lie close to the relation
defined by our original data, with the larger area surveys having higher DGRs.
Points (filled triangles)
representing the rich B-M type I clusters studied by
Lopez-Cruz et al. (1997) fall at
somewhat lower DGR than most of our clusters at similar densities. However
we should note that these clusters were selected (from a larger unpublished 
sample) {\it only} if they had LFs well fitted by a single Schechter
function. This obviously precludes clusters with steep LF turn-ups
and hence high DGR. The one comparison cluster they do show {\it with} a
turn up (A1569 at DGR $\simeq 4.2$) clearly supports our overall trend. 
Although there is now considerable scatter (and the errors on some of the points
are quite large), a weighted least squares fit to the trend gives log (DGR)
= $const. - (0.86 \pm 0.22$) log (giant density), indicating a significant
variation. 
Within the overall rich cluster
sample, if we differentiate by the clusters' B-M types there is a 
suggestion that the type Is lie at lower DGR than the others, but again this
may be biased by the Lopez-Cruz sample's selection criteria.
Finally, a steep field LF ($\alpha
\simeq -1.5$) would also give a point (filled pentagon) consistent with
the trend seen in the clusters.
 
On the other hand, Ferguson \& Sandage (1991 = FS), from a study of
fairly poor groups and clusters, deduced a trend in the opposite
direction, with the early type dwarf to giant ratio {\it increasing} for
denser clusters. However, this is not necessarily as contradictory to the
present result as it might initially appear. For instance, FS select
their dwarfs morphologically, not by luminosity (morphologically
classified dwarfs and giants significantly overlap in luminosity) and
they also concentrate solely on early type dwarfs. If, as we might expect,
low density regions have significant numbers of 
late type dwarfs (irregulars), then
the FS definition of DGR may give a lower value than ours for these regions.
Furthermore FS calculate their projected densities from {\it all} detected
galaxies, down to very faint dwarfs. Regions with high DGR will therefore
be forced to much higher densities than we would calculate for giants only.
These two effects may go much of the way to reconciling our respective
results. This is illustrated by the open triangles in figure 2 which are
an attempt to place the FS points on our system; magnitudes have
been adjusted approximately for the different wavebands, 
DGRs have been estimated
from the LFs and the cluster central 
densities (from Ferguson \& Sandage 1990)
have been scaled down by the fraction of
their overall galaxy counts which are giants (by our luminosity definition).
Given the uncertainties in the translation, 
most of the FS91 points now lie reassuringly 
close to our overall distribution.

Nevertheless, there are two exceptions, the FS points of lowest density 
(the Leo and Doradus groups) which also have low DGR (and lie close to our 
main `outlier', the point for the outer region of A22). 
The Local Group (shown by the star)
would also be in this regime, at low density and DGR = 2,
as would the `conventional' field with $\alpha \simeq -1.1$ and DGR
$\simeq 1.5$ (open pentagon). 
This may suggest that at very low density 
the trend is reversed (ie. is in the direction seen by FS), 
or that the cosmic (and/or statistical) scatter 
becomes large. More data in the very low density regime is probably
required before we can make a definitive statement on a possible reversal
of the slope of the DGR versus density relation. In particular, the scatter
in the derived faint end of the field LF between different surveys
(see, e.g., the recent discussion
in Metcalfe et al. 1998) precludes using this to tie
down the low density end of the plot. 

\section{Discussion and Summary}

As with the corresponding morphology density relation for giant galaxies, the
cause of our population - density relation could be either `nature' or
`nurture', ie. initial conditions or evolution. Some clues may be provided by
the most recent semi-analytic models of galaxy formation, which have been 
able to account in a general way for the excess of (giant) early type
galaxies in dense environments (e.g., Baugh, Cole \& Frenk 1996). 

The steep faint end slope of the LF appears to be a generic result of
hierarchical clustering models  \footnote{ And was considered a problem until
observational evidence for steep LFs increased!} 
(e.g., White \& Frenk 1991; Frenk et al. 1996;
Kauffmann, Nusser \& Steinmetz 1997),
so is naturally accounted
for in the current generation of models. The general hierarchical
formation picture envisages (mainly baryonic) galaxies forming at the cores
of dark matter halos. The halos themselves merge according to the general
Press-Schechter (1974) prescription to generate the present day halo mass
function. However the galaxies can retain their individual identities within the
growing dark halos, because of their much longer merging time scales.
The accretion of small halos by a large one then results in the main
galaxy (or cluster of galaxies, for very large mass halos) acquiring a
number of smaller satellites (or the cluster gaining additional, less 
tightly bound, members).

Kauffmann et al. (1997) have presented a detailed study of the distribution
of the luminosities of galaxies expected to be associated with a single
halo of given mass. The LFs are somewhat disjoint owing to the specific
halo masses modelled; especially for the low mass halo there is a preferred
luminosity for the central galaxy plus a tail to lower luminosities. 
For a realistic mix of halo masses, these
would no doubt be smoothed to look more like conventionally observed LFs.
Nevertheless, we can still easily compare the numbers of dwarf galaxies
per unit giant galaxy luminosity (rather than the amplitude of the giants' LF)
between halos of different mass.

The Kauffmann et al. models mimic a "Milky Way system" (halo mass $5 \times 10^{12} M_{\odot}$), a sizeable group (halo mass $5 \times 10^{13} M_{\odot}$)
and a cluster mass halo ($10^{15} M_{\odot}$). Using their figure 2 (which
also emphasises the identical faint end slopes predicted for all the
different environments), we choose to quantify the number of dwarfs by
$N_{-18}$, the number of dwarfs per system in the $M_{B} = -18$ bin. 
Because of the very similar slopes, the choice of bin or range of bins
does not affect our conclusions, so this is the equivalent of the total 
number of dwarfs used in Figure 1.
To quantify the giant population we
choose the total light of galaxies of $M_{B} = -20$ or brighter, in units
of $L_{*}$ galaxies (taking $M_{B}^{*} = -21$), which we call $N_{-21}$. 
Using this definition, rather than the actual number of galaxies brighter
than some value (as in our observational data) allows for the discretization
of the LFs for small halos. The results are summarized in Table 1. The
ratio of these two values $N_{-18}$ and $N_{-21}$ then
quantifies the relative dwarf galaxy populations. Roughly speaking, for smooth
LFs with a shape similar to that
observed, we should multiply these values by about 5, giving a range from about 
1 to 3, to compare with our observational DGRs.

\begin{table*}
\begin{center}
\begin{tabular}
{lccccccccr}
Halo Mass ($M_{\odot}$) & $N_{-18}$ & $N_{-21}$ & $N_{-18}/N_{-21}$ \\
\tableline
$5 \times 10^{12}$ & 0.2 & 0.34 & 0.6 \\
$5 \times 10^{13}$ & 2.2 & 3.8 & 0.6 \\
$1 \times 10^{15}$ & 40 & 190 & 0.2\\
\end{tabular}
\end{center}
\caption{Dwarf numbers as a function of halo mass. \label{tbl-1}}
\end{table*}

We see that the Milky Way and small group halos have similar numbers
of dwarf galaxies per unit giant galaxy light, whereas the dense cluster 
environment has a much smaller number of dwarfs for a given total giant
galaxy luminosity. Thus the predictions of the hierarchical models
(which depend, of course, on the merger history of the galaxies)
are in general agreement with our empirical results if we identify
loose clusters and the outskirts of rich clusters with a population
of (infalling?) groups (cf. Abraham et al. 1996), 
whereas the central dense regions of the clusters 
originate from already massive dark halos. By inputting realistic
star formation laws etc., Kauffmann et al. can further identify the
galaxies in the most massive halos with old elliptical galaxies, and
those in low mass halos with galaxies with continued star formation.
This would imply the likelihood that our dwarfs in low density regions
may still be star forming, or at least have had star formation in the
relatively recent past (cf. Phillipps \& Driver 1995 and references 
therein). Note, too, that these galaxy formation models would also 
indicate that the usual (giant) morphology - density relation and our 
(dwarf) population - density relation arise in basically the same way.
Finally, we can see that if these models are reasonably believable,
then we need not expect the field to be even richer in dwarfs than
loose clusters; the dwarf to giant ratio seems to level off at the
densities reached in fairly large groups.

To summarise, then,
we suggest that the current data on the relative numbers of dwarf
galaxies in different clusters and groups can be understood in terms
of a general dwarf population versus local galaxy density relation,
similar to the well known morphology - density relation for giants.
Low density environments are the preferred habitat of low luminosity
galaxies; in dense regions they occur in similar numbers to giants,
but at low densities dwarfs dominate numerically by a large factor. 
This fits in with the general idea that low luminosity
galaxies are less clustered than high luminosity ones (particularly
giant ellipticals). Plausible theoretical justifications for the 
population - density relation can be found within the context of current
semi-analytic models of hierarchical structure formation.

\clearpage
\onecolumn
\begin{figure}
\epsscale{1.0}
\plotone{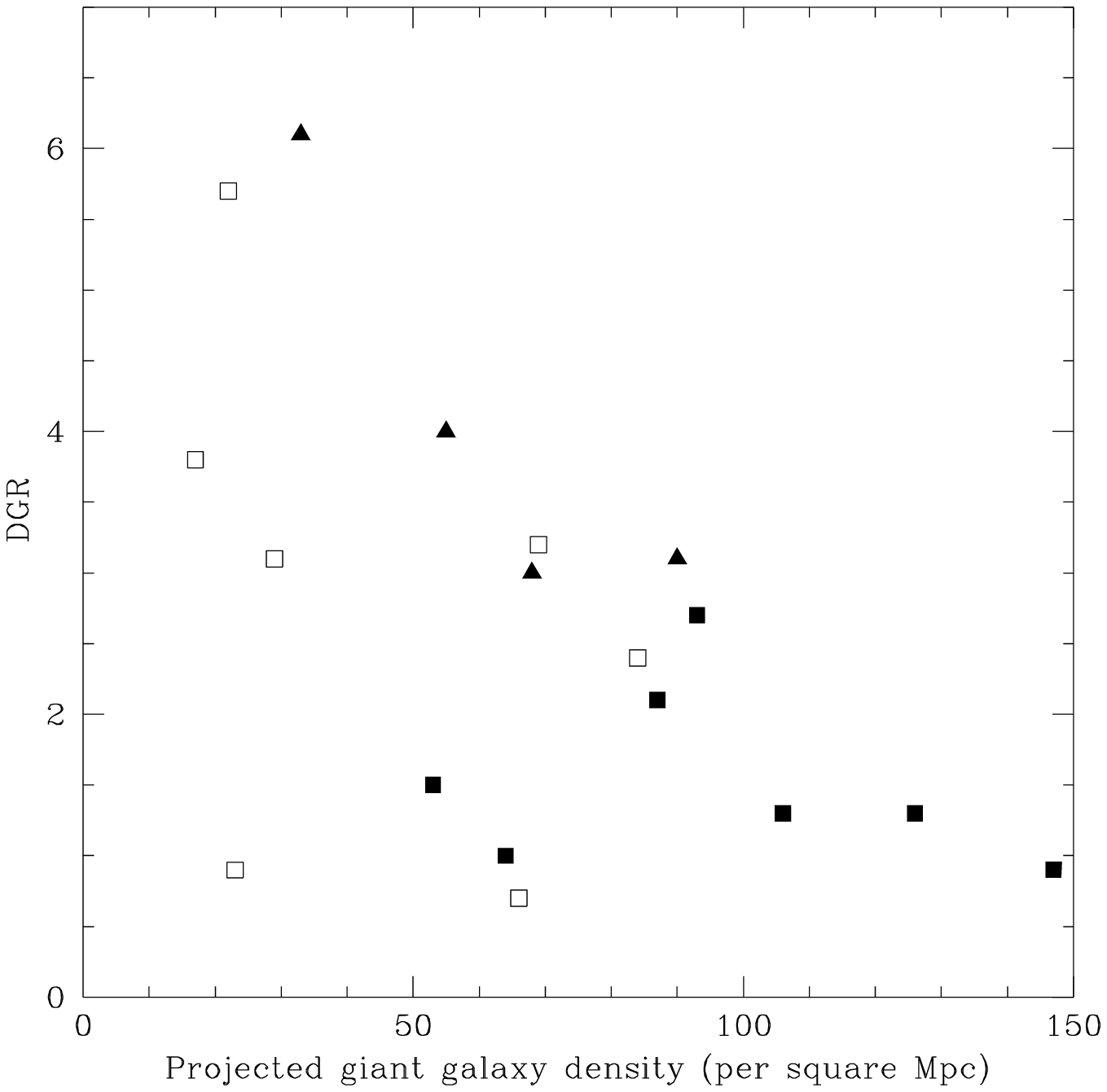}
\caption{Variation of the dwarf-to-giant ratio (DGR), as defined
in the text, with projected density of cluster giants (per square Mpc).
Solid boxes represent the central 1 square Mpc regions of the clusters,
the open boxes the outer regions (data from Paper III). 
The triangles show the variation over
a wider range of radii for Abell 2554 (data from Paper 1). Note that typical
error bars (due to the combination of poisson errors and background
subtraction errors) are 10\% in density and 20\% in DGR for the denser regions,
rising to 30\% in density and 50\% in DGR at the lowest densities (and hence
object numbers). The outlier at low density and low DGR (the outskirts
of A22) has a very large error in DGR ($\sim$ 100\%).
\label{fig1}}
\end{figure}

\clearpage
\onecolumn
\begin{figure}
\epsscale{1.0}
\plotone{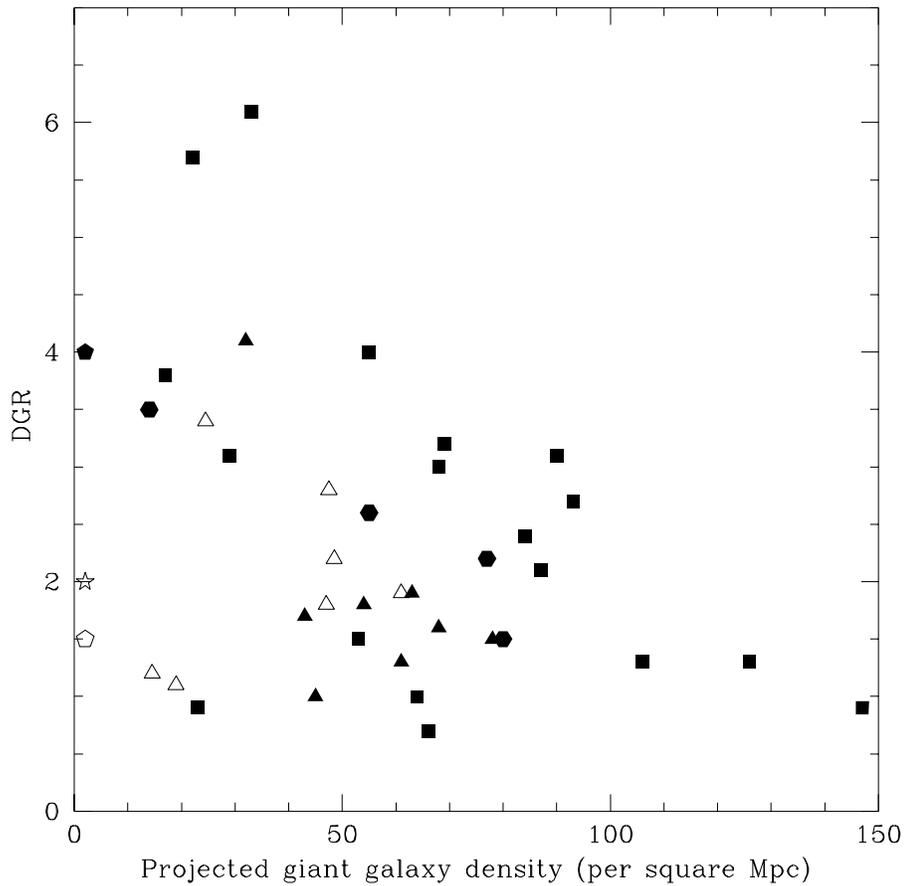}
\caption{
As figure 1, but including data from other observers. Squares are our
data repeated from figure 1, hexagons are for various Coma surveys detailed
in the text, filled triangles are from Lopez-Cruz's sample and open 
triangles are for Ferguson and Sandage's poor clusters and groups. 
The open pentagon at low density represents a conventional `flat' field LF, 
the filled pentagon a possible
steep ($\alpha \simeq -1.5$) field LF and the Local Group
is represented by the star at DGR = 2.
\label{fig2}}
\end{figure}
\end{document}